\documentclass[]{aa}  
\usepackage{graphicx}
\usepackage[authoryear]{natbib}
\bibpunct{(}{)}{;}{a}{}{,}
\usepackage{txfonts}
\begin{document}
   \title{Pre-main-sequence stars in the young open
cluster NGC 1893\fnmsep\thanks{Partially based on observations
  obtained at the 
Nordic Optical Telescope and the Isaac Newton Telescope (La Palma,
Spain) and Observatoire de Haute Provence (CNRS, France).}} 
\subtitle{II. Evidence for triggered massive star formation}
\author{I.~Negueruela
          \inst{1,2,3}
          \and
      A.~Marco\inst{1,3}
\and
G.~L.~Israel\inst{4}
\and
G.~Bernabeu\inst{1}   
	            }
   \offprints{I.~Negueruela}
   \institute{Departamento de F\'{\i}sica, Ingenier\'{\i}a de Sistemas y
  Teor\'{\i}a de la Se\~{n}al, Universidad de Alicante, Apdo. 99,
  E03080 Alicante, Spain\\
              \email{ignacio@dfists.ua.es}
\and
Observatoire de Strasbourg, 11 rue de l'Universit\'{e},
F67000 Strasbourg, France
 \and
	  Department of Physics and Astronomy, The Open University,
  Walton Hall, Milton Keynes MK7 6AA, United Kingdom 
\and
Osservatorio Astronomico di Roma, Via Frascati 33, 
I00040 Monteporzio Catone, Italy
        } 
   \date{Received }

% \abstract{}{}{}{}{} 
% 5 {} token are mandatory
 
  \abstract{The open cluster \object{NGC 1893}, illuminating the
  \ion{H}{ii} region IC~410, contains a moderately large
  population of O-type stars and is one of the youngest clusters
  observable in the optical range. It is suspected to harbour a large
  population of pre-main-sequence (PMS) stars.}
  % aims heading (mandatory)
   {We have probed the stellar population of NGC~1893 in an attempt to
  determine its size and extent. In particular, we look for signs of
  sequential star formation.} 
  % methods heading (mandatory)
   {We classify a large
  sample of cluster members with new intermediate resolution
  spectroscopy. We use H$\alpha$ slitless spectroscopy of 
  the field to search for emission line objects, identifying 18
  emission-line PMS stars. We then combine existing optical photometry
  with 2MASS $JHK_{S}$ photometry to detect stars with infrared
  excesses, finding close to 20 more PMS candidates.}
  % results heading (mandatory)
   {While almost all stars earlier than B2 indicate standard
  reddening, all later cluster members show strong deviations from
  a standard reddening law, which we interpret in terms of infrared
  excess emission. Emission-line stars and IR-excess objects show the
  same spatial distribution, concentrating around two localised
areas, the immediate vicinity of the pennant nebulae \object{Sim
129} and \object{Sim 130} and the area close to the cluster core where
the rim of the molecular cloud associated with \object{IC 410} is
  illuminated by the nearby O-type 
stars. In and around the emission nebula Sim~130 we find three
Herbig Be stars with spectral types in the B1\,--\,4 range and several
  other fainter emission-line stars. We obtain a complete census of
  B-type stars by combining Str\"omgren, Johnson and 2MASS photometry
  and find a deficit of intermediate mass stars compared to massive
  stars. We observe a relatively extended halo of massive stars
  surrounding the cluster without an accompanying population of
  intermediate-mass stars.}
  % conclusions heading (optional), leave it empty if necessary 
   {Stars in NGC~1893 show strong indications of being extremely
  young. The pennant nebula Sim~130 is an area of active massive
  star formation, displaying very good evidence for
  triggering by the presence of nearby massive stars. The overall
  picture of star formation in NGC~1893 suggests a very complex process.}


   \maketitle
%
%________________________________________________________________

\section{Introduction}

 High mass stars are known to form preferentially in star
clusters, but the exact details of how they are 
born and whether their formation has an impact on the formation of
less massive stars are still the arguments of open discussions
(see, e.g., references in Crowther 2002). As massive stars disrupt the
molecular clouds from which they are born, it is generally assumed
that the formation of massive stars closes a particular star formation
episode by eliminating the material from which further stars may form
\citep[e.g.,][]{fra94}. 
In this view, the formation of massive stars in an environment must
take place over a short timescale.

Numerous examples, however, seem to support the idea
that the presence of massive stars triggers the formation of new stars
in the areas immediately adjacent to their location
\citep[e.g.,][]{wal02}. Such scenario would explain the formation of
OB associations, extending over dozens of parsecs
\citep{el77}. Sequential star formation has been observed over both rather
small spatial scales \citep[e.g.,][]{dehar03,zav06} and  large, massive
stellar complexes, such as \object{30 Dor} \citep{wb97}, but in most
cases doubts arise about the role of the first generation of stars:
does their impact on the surrounding medium actually trigger new
star-formation episodes or simply blows away the clouds surrounding
regions where star formation was already happening anyway?

 Investigations aimed at studying these questions can
take a statistical approach \citep[as in][]{mas95} or concentrate on
the detailed study of one particular open cluster where star
formation is known to occur (e.g., the investigation of
\object{NGC~6611} by \citealt{hil93}). Unfortunately, there are not
many open clusters with active star formation and a large
population of OB stars easily accessible for these studies.

\defcitealias{chopi}{Paper~I}
\defcitealias{chopi2}{Paper~II}

\begin{figure*}[ht]
\begin{picture}(500,420)
\put(0,0){\includegraphics{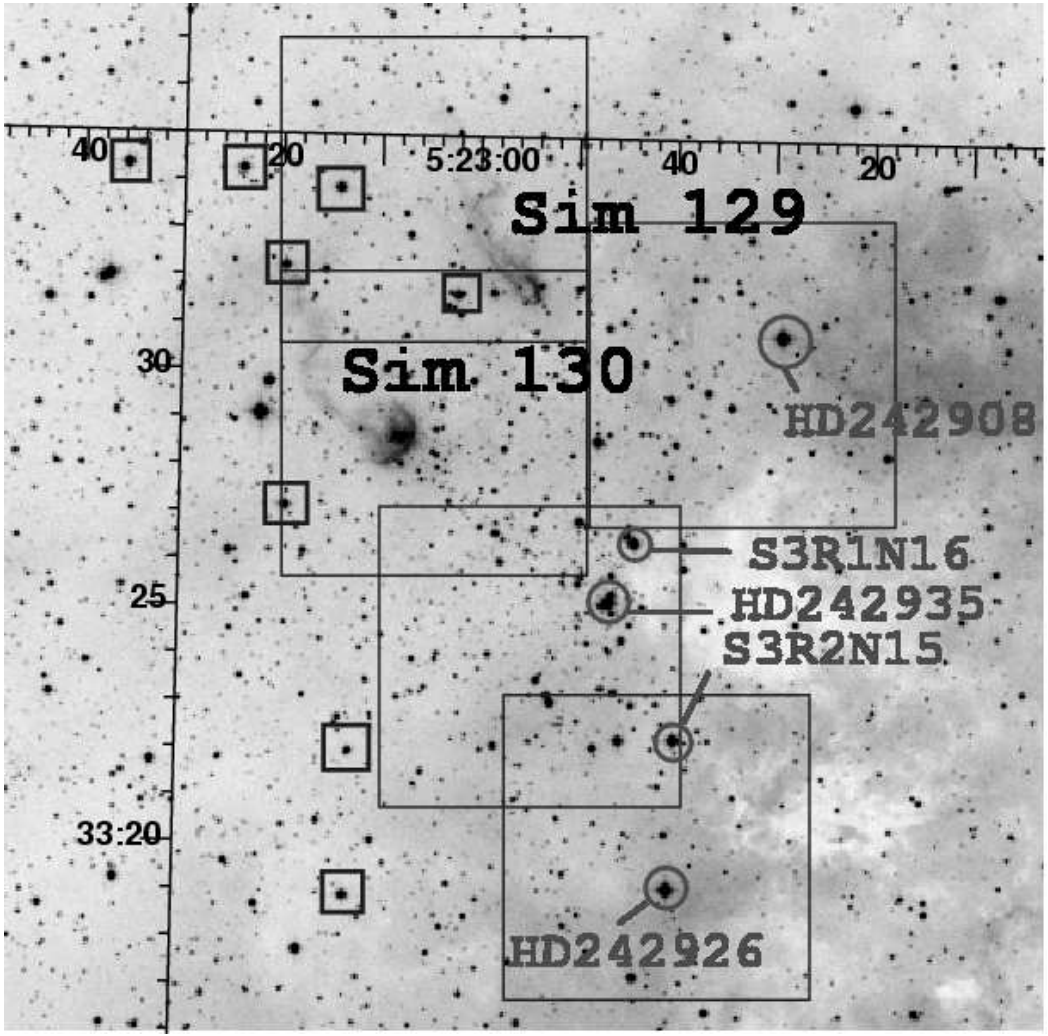}}
\end{picture}
\caption{The approximate boundaries of the five frames observed with
slitless spectroscopy are indicated on the \object{NGC~1893} field
from a digitised DSS2-red plate. Note the position of the emission
nebulae Sim~129 and Sim~130. The large white patch
almost devoid of stars to the SW is the 
molecular cloud associated with IC~410. The five O-type stars
in NGC~1893 are marked by circles and identified by name. The
8 early B-type stars in the periphery of NGC~1893 for which
we present classification spectra are identified by squares. 
The O-type stars HD~242908  and HD~242926
 are surrounded by bright nebulosity (darker patches).}
\label{fig:genview}
\end{figure*}

One such cluster is NGC 1893, a rather massive cluster, with
five catalogued O-type stars, which seems to have very recently
emerged from its parental molecular cloud. The ionising flux of the
O-type stars has generated the \ion{H}{ii} region
IC~410 on the edge of the molecular cloud (see
Fig.~\ref{fig:genview}). Though NGC~1893 is rather more distant than
some other areas where star formation can be studied, its very young
age, moderately rich O star population and relatively low interstellar
reddening make it a very interesting target for optical/infrared
studies. Moreover, its large Galactocentric distance and projection
onto a molecular cloud mean that there is very little contamination by
a background population.

%and obscured by several conspicuous dust 
%clouds, studies have generally centred on what is considered the core
%of the cluster: the area surrounding the
%O-type stars \object{S3R1N5} (O7V) and \object{S3R1N16} (O7.5V) just off
%the rim of the molecular cloud associated with \object{IC 410}. A
%clear concentration of B-type MS stars is found surrounding this two
%stars and the nearby O5.5V \object{S3R2N15} .
%More to the North, the O4V star \object{S4R2N17}
%(\object{HD 242908}), seems to be relatively separated from the
%rest of the cluster.

As with all very young open clusters, the age of NGC~1893 is
uncertain. Several authors have estimated it around $\sim 4\:$Myr
\citep{tap91,val99}, but the presence of luminosity class V early
O-type stars would indicate an age $<3\:$Myr. Moreover, some
emission-line B-type stars found in this area 
are likely to be PMS stars (\citealt{chopi2}, henceforth
\citetalias{chopi2}), and hence very young (as contraction
times for B-type stars are $<1\:$Myr). Even worse, \citet{mas95}
classify some early-B stars in NGC~1893 as luminosity class
III, implying ages $\sim 10\:$Myr. However, \citet{mas95} indicate that
their spectral classifications are relatively rough, being directed to
obtaining average spectroscopic distances rather than discussing
actual evolutionary stages. 

%A careful re-examination of the age spread
%in NGC~1893 seems thus timely.

% The distribution of  candidates was far
%from uniform, with most objects concentrating around the putative cluster
%core, but three other PMS candidates, including one emission-line
%object, were located on the Eastern edge of the area surveyed.

In \citeauthor{chopi} (\citeyear{chopi}; henceforth
\citetalias{chopi}), we used $ubvy\,{\rm H}\beta$ CCD photometry of  
$\sim 40$ very likely main-sequence (MS) members to derive $E(b-y) =
0.33\pm0.03$ and $V_{0} - M_{V} = 13.9\pm0.2$ for NGC~1893.
In \citetalias{chopi2}, we identified several PMS candidates,
based on their spectral type and observed colours, three of which were
shown to be emission-line PMS stars.

 In this paper, we investigate the possible age spread in NGC~1893 and
 take a fresh look at the star formation process in this area by
 considering a rather larger field. The paper is structured as
 follows: in Section~\ref{obs}, we present the new observations used
 in this study, which we discuss in Section~\ref{spec}, together with
 existing optical photometry. In
 Section~\ref{ir}, we use 2MASS $JHK_{S}$ photometry 
 to find stars with infrared excesses, which we identify as PMS
 candidates. We show that candidates concentrate around only two locations,
 notably in the vicinity of the bright emission nebula Sim 130. In
 Section~\ref{sim}, we present a spectroscopic study of 
 stars in the area of this nebula, identifying several emission-line
 stars. Finally, in Section~\ref{disc}, we discuss the interpretation
 of the results in terms of evidence for triggered star formation.

%__________________________________________________________________

\section{Observations and data}
\label{obs}

\subsection{New data}

We obtained imaging and slitless spectroscopy of NGC 1893
using the Andalucia Faint Object Spectrograph and Camera (ALFOSC) on
the 2.6-m Nordic Optical Telescope (NOT) in La Palma, Spain, on the
nights of December 5th-7th, 2001. The instrument was equipped with
a thinned $2048\times2048$ pixel Loral/Lesser CCD, covering a field of
view of $6\farcm4\times6\farcm4$. Standard Bessell $UBVRI$ filters
were mounted on the filter wheel, while a narrow-band H$\alpha$
filter (filter \#21, centred on $\lambda=6564$\AA\ and with
FWHM$=33$\AA) was mounted on the FASU wheel.

\begin{table}
\label{tab:spect}
\centering
\caption{Log of new spectroscopic observations. The top panel shows
  low resolution spectroscopic observations. The bottom panel displays
  intermediate resolution observations. February 2001 observations are
  from the INT. October 2001 observations were taken with the OHP
  1.93-m. December 2001 observations were taken with the NOT.} 
\begin{tabular}{lccc}
\hline\hline
Star & Date of & Dispersion & $\lambda $ Range \\ 
&observation&&\\
\hline
S1R2N38&2001 Dec 5&1.5\AA /pixel& 3830--6830 \AA\\
S2R2N43&2001 Dec 5&1.5\AA /pixel& 3830--6830 \AA\\
S1R2N26&2001 Dec 6&3.0\AA /pixel& 3100--9100 \AA\\
S1R2N26&2001 Dec 7&2.3\AA /pixel& 3100-6675 \AA\\
S5003 & 2001 Oct 25 & $1.8$\AA /pixel & 3800--6900 \AA\\
S5003&2001 Dec 6&3.0\AA /pixel& 3100--9100 \AA\\
E09&2001 Dec 6&3.0\AA /pixel& 3100--9100 \AA\\
E10&2001 Dec 7&3.0\AA /pixel& 3100--9100 \AA\\
E17&2001 Dec 7&3.0\AA /pixel& 3100--9100 \AA\\
\hline
\hline
S1R2N35 & 2001 Feb 8 & $0.4$\AA /pixel & 3950--5000 \AA \\
S1R2N14 & 2001 Oct 24 & $0.9$\AA /pixel & 3745--5575 \AA\\ 
S1R2N40 & 2001 Oct 23 & $0.9$\AA /pixel & 3745--5575 \AA\\
S1R2N44 & 2001 Oct 24 & $0.9$\AA /pixel & 3745--5575 \AA\\
S1R2N55 & 2001 Oct 22 & $0.5$\AA /pixel & 6250--7140 \AA\\
S1R2N55 & 2001 Oct 23 & $0.9$\AA /pixel & 3745--5575 \AA\\
S1R2N56 & 2001 Oct 22 & $0.5$\AA /pixel & 6250--7140 \AA\\
S1R2N56 & 2001 Oct 23 & $0.9$\AA /pixel & 3745--5575 \AA\\
S1R3N35 & 2001 Oct 23 & $0.9$\AA /pixel & 3745--5575 \AA\\
S1R3N48 & 2001 Oct 24 & $0.9$\AA /pixel & 3745--5575 \AA\\
S2R3N35 & 2001 Oct 24 & $0.9$\AA /pixel & 3745--5575 \AA\\
S2R4N3 & 2001 Oct 24 & $0.9$\AA /pixel & 3745--5575 \AA\\
S3R1N5 & 2001 Feb 9 & $0.4$\AA /pixel & 3950--5000 \AA \\
S3R1N16 & 2001 Feb 9 & $0.4$\AA /pixel & 3950--5000 \AA \\
S3R2N15 & 2001 Feb 9 & $0.4$\AA /pixel & 3950--5000 \AA \\
S4R2N17 & 2001 Feb 9 & $0.4$\AA /pixel & 3950--5000 \AA \\
Hoag 7 & 2001 Oct 23 & $0.9$\AA /pixel & 3745--5575 \AA\\
HD 243035 & 2001 Oct 24 & $0.9$\AA /pixel & 3745--5575 \AA\\
HD 243070 & 2001 Oct 24 & $0.9$\AA /pixel & 3745--5575 \AA\\
\hline
\end{tabular}
\end{table}

For the slitless spectroscopy, we
made use of the Bessell $R$ filter and grism \#4. In total, 5 slightly
overlapping images were taken, with 900-s exposure times. The weather
was relatively poor, with some thin cloud veiling present on some
of our exposures. The area covered by these observations is shown in
Fig.~\ref{fig:genview}. 
Spectroscopy of several emission-line stars in the field was taken on the same
nights using the same instrument. Grisms \#3, \#4 and \#7 were used.
%The wavelength coverages are $\lambda\lambda$ 3800$-$6800 \AA\ for grism
%\#7, $\lambda\lambda$ 3200$-$6700 \AA\ for grism \#3 and
%$\lambda\lambda$ 3200$-$9100 \AA\ for grism \#4. 
A list of the objects observed is given in Table~1.
%\ref{tab:not}.

We obtained intermediate-resolution spectra of stars in the region of
the bright nebula Sim 130 and surrounding area during 22nd-25th
October 2001, using the 1.93-m telescope at the Observatoire de
Haute Provence, France. The telescope was equipped with the long-slit
spectrograph Carelec and the EEV CCD. On the night of 22nd October, we
used the 1200 ln/mm grating in first order, which gives a nominal
dispersion of $\approx 0.45$\AA/pixel over the range
6245\,--\,7145\AA. On the nights of 23rd \& 24th October, the 600
ln/mm grating was used, giving a nominal dispersion of $\approx
0.9$\AA/pixel over the range 3745\,--\,5575\AA. Finally, on 25th
October, the 300 ln/mm grating was used, giving nominal dispersion of
$\approx 1.8$\AA/pixel over the 3600\,--\,6900\AA\ range.

Finally, intermediate-resolution spectroscopy of several bright stars
in the field of NGC 1893 
was obtained during a run at the 2.5-m Isaac Newton Telescope (INT) in
La Palma (Spain), in February 2001. Details on the configurations
used can be found in \citetalias{chopi2}, while a list of all the
observations presented here is given in Table~1.

All the data have been reduced using the {\em Starlink}
software packages {\sc ccdpack} \citep{draper} and {\sc figaro}
\citep{shortridge} 
and analysed using {\sc figaro} and {\sc dipso} \citep{howarth}. Sky
subtraction was carried out by using the POLYSKY procedure, which fits
a low-degree polynomial to points in
two regions on each side of the spectrum. The extent of these regions
and their distance to the spectrum were selected in order to reduce
the contamination due to nebular emission. Bright sky lines coming
from diffuse nebular emission are visible in almost all of our
spectra. 

\subsection{Existing data}
For the analysis, we combine the new observations with existing photometric
datasets. On one side, we use $JHK_{{\rm S}}$ photometry from the
2MASS catalogue \citep{skru06}. The completeness limit of this
catalogue is set at $K_{{\rm S}}=14.3$, which -- as we shall see --
roughly corresponds to the magnitude of
an early A-type star in NGC~1893.  In addition, we have two optical
photometric datasets. $UBV$ data from \citet{mas95} cover a wide area
around the cluster. Its magnitude limit is close to $V\sim18$, but
errors become appreciable for $V>16$. Str\"omgren photometry from
\citetalias{chopi} only reaches $V\sim16$ and covers a more restricted
central area. However, Str\"omgren photometry allows the determination
of approximate spectral types under the single assumption of standard
reddening. From \citetalias{chopi}, we know that $V\sim16$ roughly
(depending on the reddening)
corresponds to the magnitude of an early A-type star in the cluster. 
Therefore, pretty much by chance, all three datasets have comparable
limits.

\section{Results}
\label{spec}

 \subsection{Emission-line stars}

In this paper, we address the search of emission-line PMS stars
following a new approach: deep slitless spectroscopy of the whole
field. This technique, based on the use of a low dispersion grism
coupled with a broad-band filter resulting in an ``objective
prism-like'' spectrogram of all the objects in the field, has been used
by \citet{bp01} for the search of emission line stars in open clusters.
By combining a Johnson $R$ filter and a low-resolution grism, we
obtain bandpass imaging spectroscopy centred on H$\alpha$. 

The obvious advantage of this technique with respect to narrow-band
photometry (as used in \citetalias{chopi}) is that it can reach very
faint stars. The detection limit is difficult to define. In uncrowded
regions, it is reached when the spectra are too faint to be seen
against the background, but in crowded regions the overlap
of adjacent spectra becomes important. Comparison to the photometry of
\citet{mas95} shows that we 
have been able to detect emission lines in stars fainter than their
limit at $V\ga 18$, but we cannot claim completeness. 

\begin{table}
\label{tabber:new}
\centering
\caption{Known emission-line stars in NGC 1893. E01 and E02 were
  already listed in the literature. E03\,--\,E07 were described in
  \citetalias{chopi2}. E08\,--\,E18 are found or
  confirmed here. The spectral types of E09, E11 and E17 cannot be
  determined from our spectra. We did not take long-slit spectra of
  objects whose spectral type is marked as '$-$'. Note that the
  $K_{{\rm S}}$ magnitudes (all from 2MASS) of some objects are
  affected by blending.}
\begin{tabular}{lcccccc}
\hline\hline
Name& RA & Dec & Spectral&$K_{{\rm S}}$\\ 
& && Type&\\
\hline
E01 =S1R2N35&05 23 09.2& +33 30 02& B1.5\,Ve&10.01\\
E02 =S1R2N38&05 23 04.3& +33 28 46& B4\,Ve&10.25\\
E03 =\object{S3R1N3}&05 22 43.0& +33 25 05&B0.5\,IVe&10.9\\
E04 =\object{S3R1N4}&05 22 46.1&+33 24 57&B1.5\,Ve&12.3?\\
E05 =\object{S2R1N26}&05 22 48.2 & +33 25 00&$\sim$G0\,Ve&11.9?\\
E06 =\object{S2R1N16}&05 22 51.1 & +33 25 47&$\sim$F7\,Ve&11.8\\
E07 =S1R2N23 & 05 22 52.1 & +33 30 00&$\sim$F6\,Ve&11.3\\
E08 = S5003 & 05 22 40.8& +33 24 39 &Ke&13.5 \\
E09 & 05 22 43.8&+33 25 26&?& 9.4\\
E10 & 05 22 49.6&+33 30 00&?& $>$15\\
E11 = \object{S1R2N26} & 05 22 57.9 & +33 30 42&$\sim$A3\,Ve&12.6\\
E12 & 05 23 00.0 &+ 33 30 41&?&13.0\\
E13 & 05 23 02.8 &+ 33 29 40&$-$&13.9\\
E14 & 05 23 04.4 &+ 33 29 48&$-$&13.2\\
E15 & 05 23 06.3 & +33 31 02&$-$&13.2\\
E16 = S1R2N55N & 05 23 08.3 & +33 28 38&B1.5\,Ve&$>10.3$\\
E17 & 05 23 08.9 &+ 33 28 32&?&11.5\\
E18 & 05 23 09.9 &+ 33 29 09&$-$&13.2\\
\hline
\end{tabular}
\end{table}

We detect the 7 previously known
emission-line objects (the 5 listed in \citetalias{chopi2} and the two
catalogued H$\alpha$ emitters close to Sim 130). 
We fail to detect the candidate emission-line \object{S5003}
\citepalias{chopi2}, as its spectrum, given 
the orientation of our dispersion direction, is completely superimposed
by that of its brighter neighbour \object{S3R1N9}. However, we have
obtained a long-slit spectrum of this object and can confirm it as an
emission-line star (see Table~\ref{tabber:new}). In addition, we detect 10
new stars with H$\alpha$ emission (see Table~\ref{tabber:new}), and name them
E09\,--\,E18. Only two of the new emission-line objects are bright
enough to have been observed by previous authors, namely
\object{S1R2N55} and \object{S1R2N26}. We confirmed the emission-line
nature of some of these objects through long-slit spectroscopy
(Table~1).
%\ref{tab:not}
Most of them are too faint for spectral
classification. S1R2N26 is an A-type star with strong H$\alpha$
emission. E12 does not show any photospheric features, but has all
Balmer lines in emission and also shows strong \ion{Ca}{ii} K lines and
\ion{Ca}{ii} 8498, 8542,8662~\AA\ triplet emission. All the known
emission-line stars in NGC~1893 are listed in Table~\ref{tabber:new}.

\subsection{O-type stars}

We have obtained accurate classifications for four O-type stars in
the field of NGC 1893. The fifth O-type star, to the South of
the cluster (\object{HD 242926}), was not observed here, but it has been
observed as part of another programme and its spectral type is O7\,V,
in total agreement with \citet{wal73}. This star is reported to show
strong radial velocity changes by \citet{jon72}.

\begin{figure}
\begin{picture}(250,180)
\put(0,0){\includegraphics{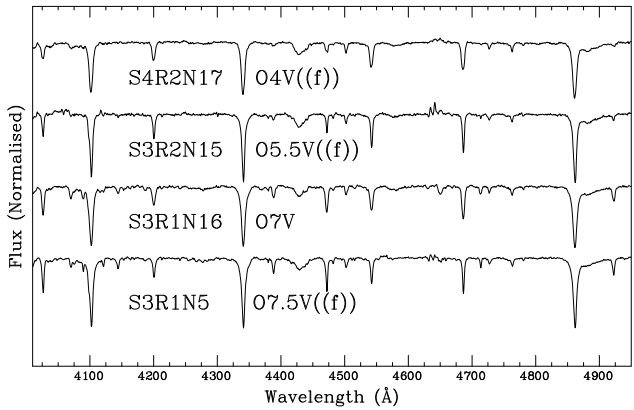}}
\end{picture}
\caption{Classification spectra of the 4 O-type stars near the core of
NGC 1893. The earliest spectral type is that of
S4R2N17 (\object{HD 242908}), O4\,V((f)). The other three
objects have spectral types O5.5\,V((f)) for S3R2N15, O7\,V for
S3R1N16 and O7.5\,V((f)) for S3R1N5.}
\label{fig:photonpower}
\end{figure}

The spectra of the four O-type stars observed are presented in
Fig.~\ref{fig:photonpower}. The spectral types have been derived using
the quantitative methodology of \citet{mat88}, based on Conti's
scheme. For \object{S4R2N17} (HD~242908), the spectral type criteria
fall on the 
border between O4 and O5 and we will adopt the O4\,V((f))
classification given by \citet{wal73}. 
In \object{S3R2N15} (LS~V +34$\degr$15), the \ion{He}{i} lines are
stronger and the 
quantitative criteria indicate O5.5\,V((f)). \citet{jon72} reports two
measurements of the radial velocity of this star, differing by more
than $100\:{\rm km}\,{\rm s}^{-1}$, strongly suggesting that there are
two O-type stars in this system. In \object{S3R1N16}
(BD~+33$\degr$1025A), the condition \ion{He}{i}~$\lambda$4471\AA\ $\simeq$\
\ion{He}{ii}~$\lambda$4541\AA\ implies by definition O7, while the
strength of \ion{He}{ii}~$\lambda$4686\AA\ and very weak \ion{N}{iii}
emission indicate a MS classification. Finally, in \object{S3R1N5}
(HD~242935), \ion{He}{i}~$\lambda$4471\AA\ is slightly stronger than
\ion{He}{ii}~$\lambda$4541\AA\ and the quantitative criteria indicate
O7.5\,V((f)), though \ion{He}{ii}~$\lambda$4686\AA\ is rather weak and
close to the limit for luminosity class III given by \citet{mat88}.

\subsection{B-type stars}

\begin{figure}[ht]
\resizebox{\columnwidth}{!}{\includegraphics[bb = 80 110 490 700]{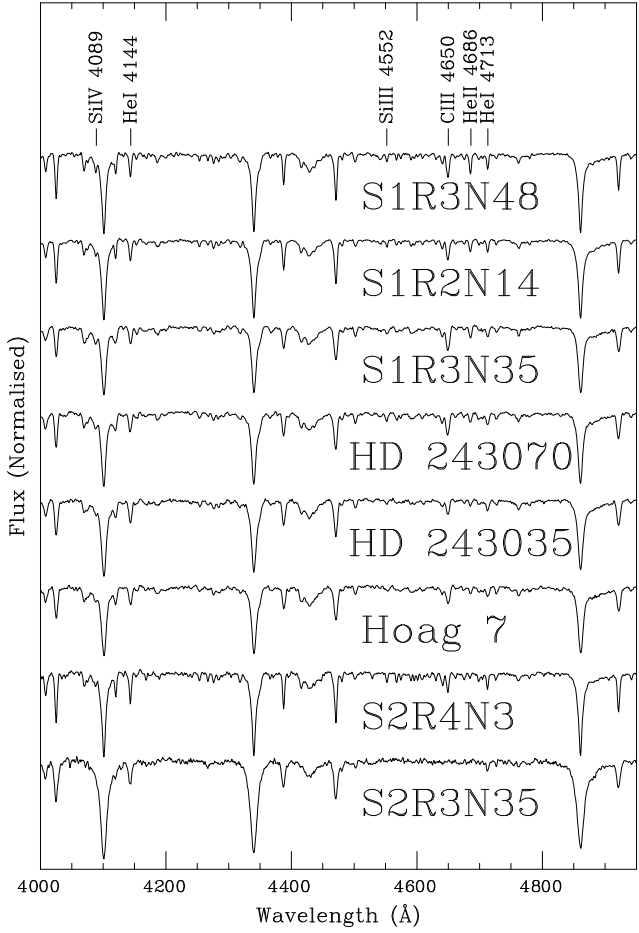}}
\caption{Classification spectra of B-type stars in the area of
NGC 1893. From top to bottom, stars are displayed 
from earlier to later spectral type. Photospheric features used
for spectral classification in this range are indicated.}
\label{fig:bstars}
\end{figure}

We have also derived new classifications for bright B-type stars to
the East of NGC 1893, in order to check if they are connected
to the cluster in spite of their relatively large angular distance to
the cluster core (see their distribution in
Fig.~\ref{fig:genview}). These spectra are displayed in
Fig.~\ref{fig:bstars}.  
At the resolution of our spectra, the traditional criterion for
spectral classification  
around B0, namely the ratio between  \ion{Si}{iv}~$\lambda$4089\AA\ and 
\ion{Si}{iii}~$\lambda$4552\AA\ \citep{waf} is difficult to apply, since 
\ion{Si}{iv}~$\lambda$4089\AA\ is blended into the blue wing of H$\delta$. 
We have thus resorted to additional criteria. 

\begin{table}
\label{tab:types}
\centering
\caption{O-type and early B-type stars around NGC~1893 for which we
  derive accurate spectral types, together with the derived reddening.
Photometric data are from \citet{mas95}. For HD~242935 there are
  no reliable photometric measurements due to heavy blending.}
\begin{tabular}{lcccc}
\hline\hline
Star & Name&$V$ & Spectral& $E(B-V)$ \\ 
Number&& &Type&\\
\hline
S4R2N17&HD 242908& 9.03 & O4\,V((f))&0.57\\
S3R2N15&LS V +34\degr15& 10.17 &O5.5\,V((f))&0.79\\
S3R1N16& BD +33\degr1025A&10.38&O7\,V&0.59\\
S3R1N5& HD 242935 &$-$&O7.5\,V((f))& $-$\\
\hline
\object{S1R2N14} & \object{LS V +33\degr23}& 11.22 &B0.2\,V &0.45\\
\object{S1R3N35} & \object{LS V +33\degr26}& 11.17 &B0.2\,V &0.54\\
S1R3N48 & \object{HD 243018}& 10.94 &B0\,V &0.42\\
\object{S2R3N35} & $-$ &12.43 & B2.5\,V&0.47\\
\object{S2R4N3} & \object{LS V +33\degr24} & 11.02 &B0.5\,V&0.66 \\
&\object{HD 243035} & 10.90 & B0.3\,V &0.50\\
&\object{HD 243070} & 10.83& B0.2\,V &0.51\\
\object{Hoag 7} & \object{LS V +33\degr27}& 10.74 &B0.3\,V &0.49\\
\hline
\end{tabular}
\end{table}

\object{S1R3N48} is the earliest object in the sample, and the only one 
for which we give a B0\,V classification, based on the conditions 
\ion{He}{ii}~$\lambda$4686\AA\,$\simeq$\,\ion{C}{iii}~$\lambda$4650\AA\
and  \ion{He}{ii}~$\lambda$4686\AA$\,>\,$\ion{He}{i}~$\lambda$4713\AA.
We have classified as B0.2\,V those stars in which
\ion{He}{ii}~$\lambda$4200\AA\ is barely visible and
\ion{He}{ii}~$\lambda$4686\AA\ is comparable in strength to
\ion{He}{i}~$\lambda$4713\AA. We have taken as B0.3\,V those stars in
which  \ion{He}{ii}~$\lambda$4200\AA\ is not visible and
\ion{He}{ii}~$\lambda$4686\AA\ is clearly weaker than
\ion{He}{i}~$\lambda$4713\AA, and assigned B0.5\,V to those objects in
which \ion{He}{ii}~$\lambda$4686\AA\ is visible, but very weak. All
the stars analysed fall in the B0\,--\,B0.5 range (see
Table~3), except for  
\object{S2R3N35}. For this object, we derive a spectral type B2.5\,V,
%based on the strength of \ion{Si}{ii}~$\lambda$4128\AA\ and 
%\ion{Mg}{ii}~$\lambda$4481\AA.
In this spectral range,
the luminosity class is mainly indicated by the strength of the
\ion{Si}{iii} and \ion{Si}{iv} lines compared to those of \ion{He}{i},
while \ion{C}{iii} and \ion{O}{ii} lines can be used as subsidiary
indicators \citep[see][ for a detailed discussion]{waf}. None of the
spectra analysed suggests a luminosity class different from V. 

For all the B-type stars with accurate spectral types either here or in
\citetalias{chopi}, we calculate the distance modulus, using $UBV$
magnitudes from \citet{mas95}, intrinsic 
colours of \citet{weg94} and absolute magnitudes from
\citet{hme84}, under the assumption of standard reddening (justified
in the next section). The average value is $13.4\pm0.3$, in good agreement
with the value found by \citet{mas95}.

\section{2MASS data}
\label{ir}

\subsection{Interstellar Reddening}

The reddening along the face of NGC~1893 is known to be variable and
this may have a bearing on the derivation of cluster parameters, as
most methods are very sensitive to the interstellar reddening law
assumed and treatment of the reddening. Deviations from the standard value are 
frequent in the optical, though the reddening law in the infrared has
been proved to show very little variability along different lines of
sight \citep{inde05}. Because of this, we use the 2MASS $JHK_{{\rm S}}$
photometry to check if the extinction law towards
NGC~1893 is standard. 

For all B-type member stars with accurate Str\"omgren photometry in
\citetalias{chopi}, we calculate individual values of
$E(B-V)$ by using the simple relation $E(B-V)=1.4E(b-y)$. From the
2MASS magnitudes, 
we calculate $E(J-K_{\rm S})$ assuming the intrinsic colours from
\citet{ducati} and the spectral types derived from the Str\"omgren
photometry. For a standard reddening law
\citep{rl85}, we should have $E(J-K_{\rm S})\simeq 0.5
E(B-V)$\fnmsep\footnote{The extinction law of \citet{rl85} uses 
  Johnson's $K$ instead of $K_{{\rm S}}$, but the difference is
  negligible. Using the values calculated by \citet{han03}, we would
  have $E(J-K_{{\rm S}})= 0.48 E(B-V)$}. We find that many stars have 
$E(J-K_{\rm S})\approx 0.5 E(B-V)$, but a significant
fraction show rather larger $E(J-K_{{\rm S}})$ than
expected from their $E(B-V)$ .

{\begin{figure}
\begin{picture}(500,240)
\put(0,0){\includegraphics{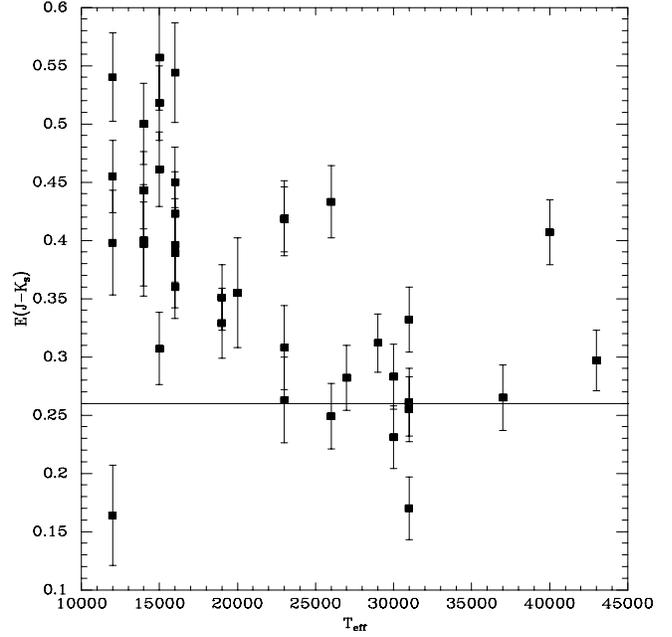}}
\end{picture}
\caption{Plot of the infrared excess $E(J-K_{\rm S})$ against spectral
  type (represented by the corresponding $T_{\rm eff}$) for O and
  B-type MS stars in the cluster. The straight line $E(J-K_{\rm
  S})=0.26$ corresponds to the cluster average $E(B-V)=0.53$
  \citep{mas95} if a standard reddening is assumed. The error
  bars in $E(J-K_{\rm S})$ represent only the photometric errors. The
  uncertainty in the intrinsic colours has not been included, as it is
  difficult to quantify and unlikely to depend on spectral type. }  
\label{fig:teff}
\end{figure}

This deviation from the standard relationship is very unlikely due to
a non-standard extinction law, because the amount of
excess $E(J-K_{\rm S})$ is not strongly 
correlated with position within cluster. On the other hand, 
Fig.~\ref{fig:teff} shows the dependence of $E(J-K_{\rm S})$
with spectral type. For all O and B-type members, we plot $E(J-K_{\rm
  S})$ against the $T_{\rm eff}$ corresponding to the spectral type
derived according to the 
calibrations of \citet{mar05} for O-type stars and \citet{hme84} for
B-type stars. Except for three stars, all stars earlier than
B3 have about the same $E(J-K_{\rm S})$, compatible or very slightly
above the value 
expected for a standard reddening law\fnmsep\footnote{Note that
  S3R2N15 has a high $E(J-K_{{\rm S}})$, but it also has $E(B-V)$ much
  above other stars in Table~3.}. All stars with
spectral type B3 or later (except for two) have larger excesses, with
a clear tendency to have larger excesses as we move to later
spectral types. This dependence of the 
infrared excess on spectral type, while it shows no dependence on
location, is clearly suggesting that the excesses are intrinsic to the
stars and not related to the extinction law.

To investigate this further, we use the {\sc chorizos} code
\citep{maiz} to 
estimate the extinction law. This program tries to reproduce an
observed energy spectral distribution by fitting
extinction laws from \citet{cardelli} to the spectral distribution of
a stellar model. As input, we used the $UBV$ photometry from
\citet{mas95} and the $JHK_{{\rm S}}$ photometry from 2MASS, together
with Kurucz models of main sequence stars with the 
 $T_{{\rm eff}}$ corresponding to our stars and $\log g =4.0$. We run
the program for all the stars with spectral types derived from spectra
or from Str\"omgren photometry. Out of 20 stars earlier than B3, 15 are
best fit by reddening laws having $2.8<R<3.4$, while 5 require
$R>3.5$. Out of 28 stars later than B3, 25 require $R\geq 3.5$ and 3 did
not converge to a solution. Again, we find a strong dependence of the
reddening law on the spectral type. Taking the average for all the
stars with spectral type earlier than B3, we find
$R=3.3\pm0.2$. Leaving out the 5 stars with $R>3.5$, we find an
average $R=3.16\pm0.12$.

\begin{figure}
\begin{picture}(500,240)
\put(0,0){\includegraphics{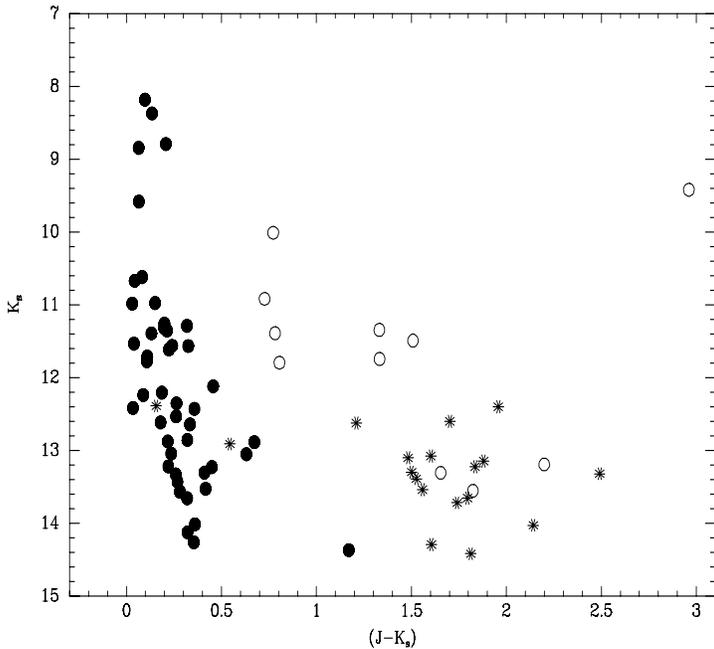}}
\end{picture}
\caption{Plot of $K_{\rm S}$ against $(J-K_{\rm S})$ for all the 2MASS
  stars fulfilling the two conditions set in
  Section~\ref{sec:2mass}. Filled circles 
  represent O and B-type MS members from \citet{chopi} or with spectra
  in this paper. Open circles are
  known emission-line PMS stars. Stars represent the rest of the IR excess
  candidates, three of which fall within the
  main sequence traced by known members (one has a known member
  superimposed).
 Note also the object with $(J-K_{{\rm
  S}})\approx3$, our emission-line source E09.} 
\label{fig:(J-K)-K}
\end{figure}

Our interpretation of this result is that the interstellar reddening
law to NGC~1893 is standard, but a substantial number of objects show
important individual $(J-K_{\rm S})$ excesses. These excesses are
present in all stars later than B3 and in a few early stars (these
early-type stars with anomalous values of $R$ may perhaps 
have later-type companions with individual excesses).
This interpretation is further supported by the 
$K_{\rm S}$/$(J-K_{\rm S})$ diagram for 
cluster members (Fig.~\ref{fig:(J-K)-K}). All stars of mid and
late B spectral type deviate strongly from the almost vertical main
sequence traced by early members, displaying much larger $(J-K_{\rm
  S})$. This separation is in clear contrast with the fact that stars
in the B0\,--\,B8 range have almost identical $(J-K)_{0}$
\citep{ducati} and again confirms that all stars later than B3 show
some $E(J-K_{{\rm S}})$ excess, with stronger excesses corresponding
to later spectral types. 

Mid and late B stars fit the ZAMS rather well in different optical
observational HR diagrams \citep[e.g.,][]{tap91,mas95,chopi}, though
they do not occupy standard positions in the $[c_{1}]/[m_{1}]$ diagram
\citep{tap91,chopi}. How do we then
explain their $E(J-K_{{\rm S}})$ excesses? In Fig.~\ref{fig:excess},
we plot an infrared HR diagram for the B-type members. For each star,
we assume that the interstellar reddening $E(J-K_{{\rm S}})_{\rm is} =
0.5 E(B-V)$ and the corresponding extinction is $A_{K_{{\rm
      S}}}=0.67E(J-K_{{\rm S}})_{\rm is}$, according to the standard
reddening law. We then 
shift their $m_{K_{{\rm S}}} = K_{{\rm S}}-A_{K_{{\rm S}}}$ by DM$ =
13.5$ (see Section~\ref{sec:dist}) and plot them in the
colour-magnitude diagram. We also plot the PMS isochrones from
\citet{siess} for 1 and 2 Myr, together with the youngest
isochrone from \citet{gir02}, corresponding to 4~Myr, as the
position of most B stars on this isochrone will not deviate in any
measurable way from the ZAMS.

\begin{figure}
\begin{picture}(500,240)
\put(0,0){\includegraphics{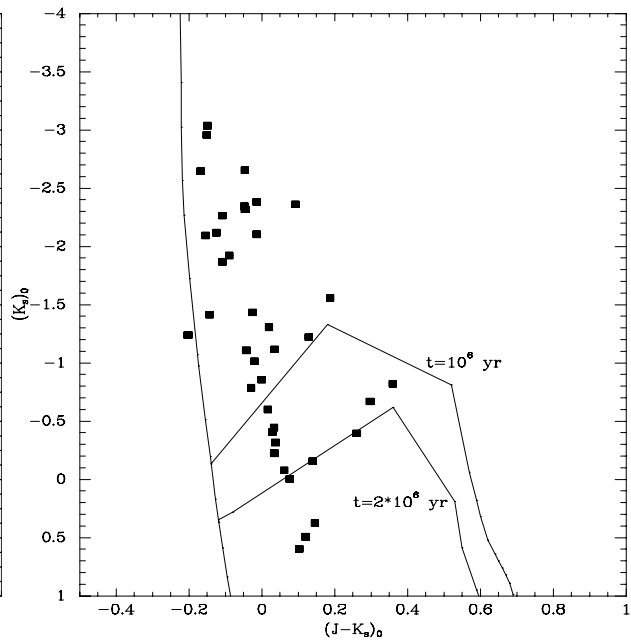}}
\end{picture}
\caption{Observational HR diagram for B-type members of NGC~1893. The
  location of the ZAMS is represented by the 4-Myr isochrone from
  \citet{gir02}. The PMS isochrones for 1 and 2 Myr from
  \citet{siess} are also shown.}   
\label{fig:excess}
\end{figure}

We observe that the smooth way in which the $E(J-K_{\rm S})$ excesses
increase with decreasing mass does not fit at all the shape of the PMS
isochrones. As a matter of fact, the positions of the stars in the HR
diagram cannot mean that they are moving towards the main sequence
along PMS tracks, because the infrared excess, as it has been defined,
is a measure of the discrepancy between optical and infrared
colours. PMS stars should have both optical and infrared colours
appropriate for the position in the theoretical HR diagram.
Indeed, if the deviation of B3\,--\,4 stars from the ZAMS had
to be attributed to their being on PMS tracks, the age of the
cluster should be only a few $10^5$~yr. In view of this and
the fact that most B-type stars fit the ZAMS well    
in the optical, we are led to interpret the $E(J-K_{\rm S})$
excesses as due to the
presence of material left over from the star formation process, most
likely in the form of a remnant disk. If so, the PMS isochrones would
suggest an age for the cluster of $\sim 2$~Myr, with most stars later
than B2 (i.e., stars with $M_{*}<8M_{\sun}$) showing evidence for some
remnant of a disk. This remnant cannot be very important, because
  stars fit the ZAMS well in the optical (both Johnson and
  Str\"omgren) HR diagrams and are not detected as emission line
  objects. High SNR spectra of the H$\alpha$ line or in the $H$ or $K$
  bands may be able to reveal some spectroscopic signature of such remnants
  and so confirm this hypothesis.

\subsection{OB stars and PMS selection}
\label{sec:2mass}

The OB stars observed extend over a wide region to the East of the
molecular cloud associated with IC~410. The core of NGC~1893 (defined
as the region with the highest concentration of MS members) lies
immediately to the West of the molecular cloud. There is an abrupt
decrease in the number of optically visible stars as we move East from
the O-type stars S3R1N16, HD 242935 and S3R2N15 (see
Fig.~\ref{fig:genview}). This suggests that an opaque part of the
molecular cloud blocks our view \citep[see also][]{lei89}. As we lack
three dimensional 
information, we cannot know if the cluster is spread over the wall of
the dark cloud, but the fact that there is no bright luminosity in
this area suggests that the dark cloud is partially between us and the
cluster. As large areas of dark
nebulosity block some sight-lines, we consider the possibility that there
may be other OB stars in the field, obscured by gas and dust.
  
The $J$, $H$ and $K_{\rm S}$ magnitudes from the 2MASS catalogue can
be used to look for 
reddened early-type stars, under the assumption of a standard reddening
law. Taking $M_{K}=-1.6$ as the intrinsic
magnitude  for a B2 V star \citep{hme84,ducati} and $DM=13.9$ to
NGC~1893 (considered an upper limit), all stars earlier than B2
located at the distance of the 
cluster and reddened according to the law of \citet{rl85} 
will fulfil the following condition:
\begin{equation}
K_{\rm S}-1.78(H-K_{\rm S})<12.5
\label{eq1}
\end{equation}

Obviously many other stars will also fulfil this condition. However,
if we define the reddening free parameter $Q=(J-H)-1.70(H-K_{S})$, OB
stars will have $Q\simeq0.0$.
Combination of Condition~\ref{eq1} with $Q\simeq0.0$ has been shown to be
very efficient at identifying reddened OB stars
\citep[e.g.,][]{cp05}. However, not only OB stars fulfil both
conditions. Foreground A-type stars and background red giants and
supergiants will also fulfil the conditions. However, in our case,
contamination by background red stars is unlikely, as the cluster lies
at a large Galactocentric distance in the Anticentre direction and is
projected on to a dark cloud.

We take all 2MASS objects within a $15\arcmin$ radius of the centre of
the cluster with magnitudes tagged as good and an error in the $K_{\rm
  S}$ magnitude $\delta K_{\rm S}\leq 0.05$ mag (larger errors would
imply completely unreliable $Q$ parameters) and select all stars
fulfilling Condition~\ref{eq1} and having $Q<0.1$.
We discard foreground A-type stars using two criteria: (1) $(B-K_{\rm
  S})\approx0$ indicates unreddened stars (most stars have $B$
  magnitudes from \citet{mas95}; for 
  the rest we use USNO B1.0 magnitudes), and (2) $(J-K_{\rm S})$ lower 
than the average for the known OB members indicates that the stars
  are foreground to the cluster (only objects substantially more
  reddened than this average would have escaped optical surveys).

All known members earlier than B2 are selected by these two
criteria. Moreover, because of their infrared excesses, several B-type
members later than B2 are also selected. As a matter of fact, a
substantial fraction of the emission-line stars listed 
in Table~3 (the brightest ones), including some of F type,
have been selected. This is not surprising as these stars have strong
infrared excesses and then : (a) are much
brighter in  $K_{{\rm S}}$ than normal stars of the same
spectral type (so much brighter that they pass our magnitude cut) and
(b) the assumption of a standard interstellar law in the calculation
of the $Q$ parameter means that their
$(H-K_{\rm S})$ excesses are ``over-corrected'', resulting  in
negative values of $Q$, values that no normal star can have.

In view of this, we take stars fulfilling Condition~\ref{eq1}, having
$Q<-0.05$ and large values of $(J-K_{{\rm S}})$ as infrared excess
objects and therefore PMS star candidates. This procedure does not
select all the objects with infrared excess in the field, but only
those relatively bright and with strong $(H-K_{{\rm S}})$ excesses. A
search for 
a complete sample of infrared excess objects in this field would need
deeper photometry than provided by 2MASS and is beyond the scope of
this paper.

\begin{figure*}[ht]
\begin{picture}(500,420)
\put(0,0){\includegraphics{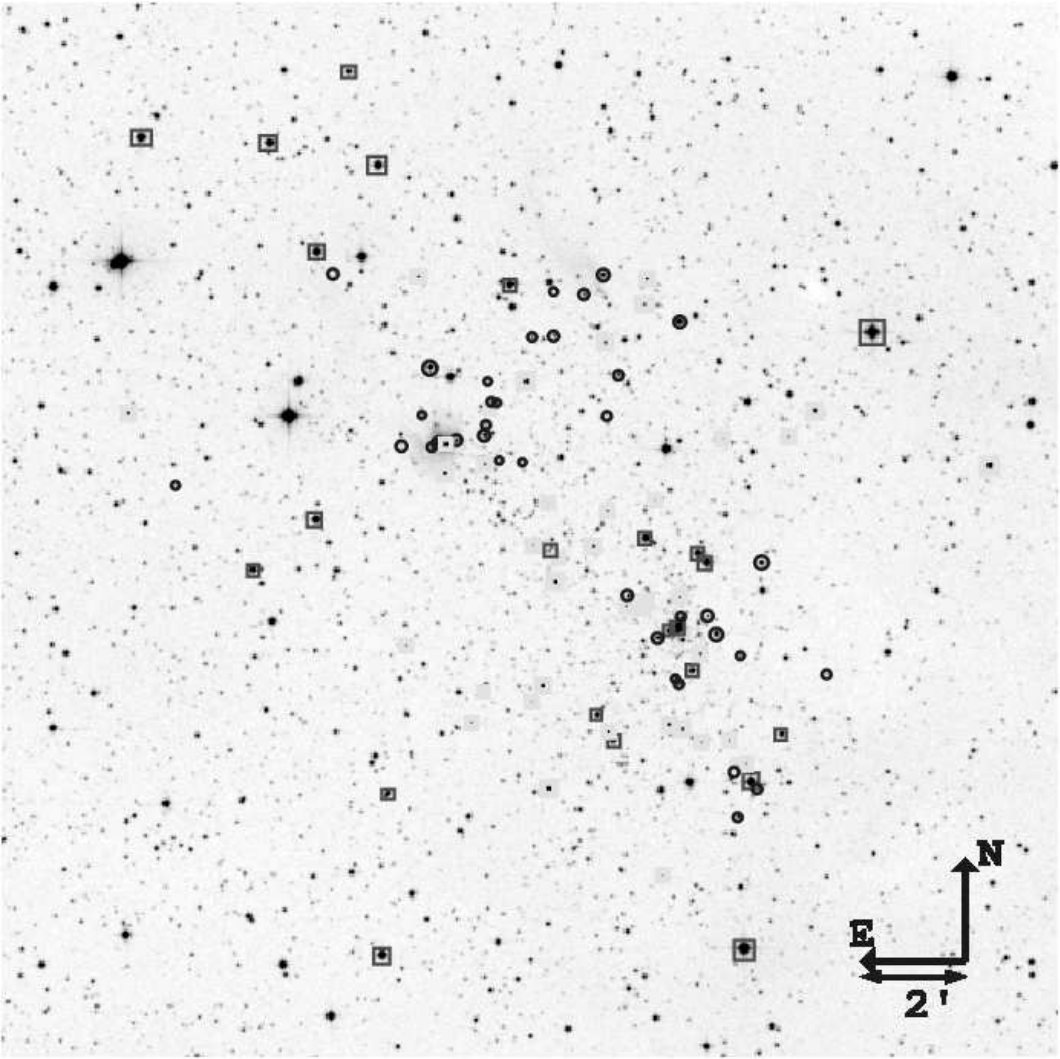}}
\end{picture}
\caption{The spatial distribution of all known members of NGC~1893 is
  shown on this DSS2 image. Objects on the main sequence according to
  their optical colours are shown as
  squares (red for spectral types B2 and earlier, yellow for B3 and
  later). PMS candidates (emission line stars and infrared excess
  objects) are shown as blue circles. Main-sequence objects are
  spread over a much larger area than PMS candidates.}
\label{fig:ms_pmsdistribution}
\end{figure*}

We find more than thirty stars fulfilling those criteria, among them,
11 of the known
emission-line stars. Figure \ref{fig:(J-K)-K} shows the $K_{\rm
  S}/(J-K_{\rm S})$ 
diagram for all MS members (filled 
circles) in NGC 1893 selected in \citet{chopi} and all the infrared
excess stars. Known emission-line PMS
stars (open circles) are located to the right of the main sequence, as
expected. We see that all the newly selected candidates
occupy the same locations as the emission-line stars, but at fainter
magnitudes, except for three objects falling close to the
  location of B-type MS
members. Of these, one is a photometric member outside the area
covered in \citetalias{chopi}, according to its $UBV$ magnitudes. A
second one is very far away from the
cluster and could only be a member if it is much more reddened than
any other member. Hence we reject it as a good photometric
candidate. The third object is S3R1N13, which was identified in 
\citetalias{chopi2} as a candidate PMS star without emission lines. It
has a spectral type B5\,III-IV and its high reddening strongly
suggests it is a PMS star in NGC~1893. The only alternative would be a
foreground star with very unusual colours, but its spectrum does not
show significant anomalies.

Fig.~\ref{fig:ms_pmsdistribution} compares the distribution of
these candidate infrared excess objects to that of MS members.
The distribution is certainly not
random, as they strongly concentrate around two small areas, the
vicinity of the pennant nebulae Sim~129 and Sim~130 and the rim of the
molecular cloud closest to the cluster core. This
distribution confirms beyond doubt that the stars selected are a
population associated with the cluster rather than red background
stars. Moreover, the spatial distribution of these objects coincides
exactly with that of emission-line stars, giving full support to the
interpretation that most of them are also PMS stars. We cannot rule out
the possibility that a few of these objects (especially the few ones at large
distances from the cluster) are background red stars, but certainly
the majority of these objects are PMS members with large infrared
excesses.

Once we account for known members, foreground stars, emission-line
stars and infrared-excess candidate PMS stars, we have exhausted the
list of stars selected according to our original criteria. This means
that there are no obscured OB 
stars in this region, at least within the magnitude limit of
2MASS. Note, however, the position of E09 in
Fig.~\ref{fig:(J-K)-K}. If this object belongs to the cluster, it
should be a young massive stellar object. Unfortunately, this object
is so faint in the optical that our spectrum is extremely noisy. The
only obvious feature is a very strong H$\alpha$ emission line. Another
possible emission feature may correspond to the \ion{O}{i}~8446\AA\ line.

\section{The area around Sim 130}
\label{sim}

Results in the previous sections clearly show that present-day star
formation in NGC~1893 is strongly concentrated towards the pennant
nebulae Sim~129 and Sim~130, with most of the PMS stars located
around the latter.
  
The ``head'' of  Sim 130 contains a group of stars which were
observed photometrically by \citet{tap91} as if they were a
single object (their Star 35). They derive the colours of an
early-type star with emission lines. The head of Sim 130 has
also been identified as the near-infrared counterpart of the IRAS source
\object{IRAS 05198+3325}, considered a Young Stellar Object (YSO)
candidate (CPM16 in \citealt{cpm}). In addition, in the
immediate vicinity of Sim 130, we find the emission-line star
\object{S1R2N35}, which based on low-resolution spectra, was
considered to be a very good candidate to a Herbig Be star in
\citetalias{chopi}. Not very far away lies the catalogued emission-line B star
\object{NX Aur} = \object{S1R2N38}.

Figure~\ref{fig:closeup} identifies the brightest stars
immersed in the bright nebulosity of Sim 130:
\object{S1R2N56} (\object{[MJD95] J052307.57+332837.9} in the
catalogue of \citealt{mas95}), \object{S1R2N55} (\object{[MJD95]
  J052308.30+332837.5}) and a fainter star not observed by
\citet{cuf73}, which we will 
call \object{S5004} (\object{[MJD95] J052306.71+332840.3}). There are
many other fainter stars within the nebula (more clearly seen in
$R$-band images), among them our emission-line objects E17 and E18.
Also, partially immersed in the nebulosity, we find 
the bright star \object{S1R2N44}.

\begin{figure}
\resizebox{\hsize}{!}{\includegraphics[]{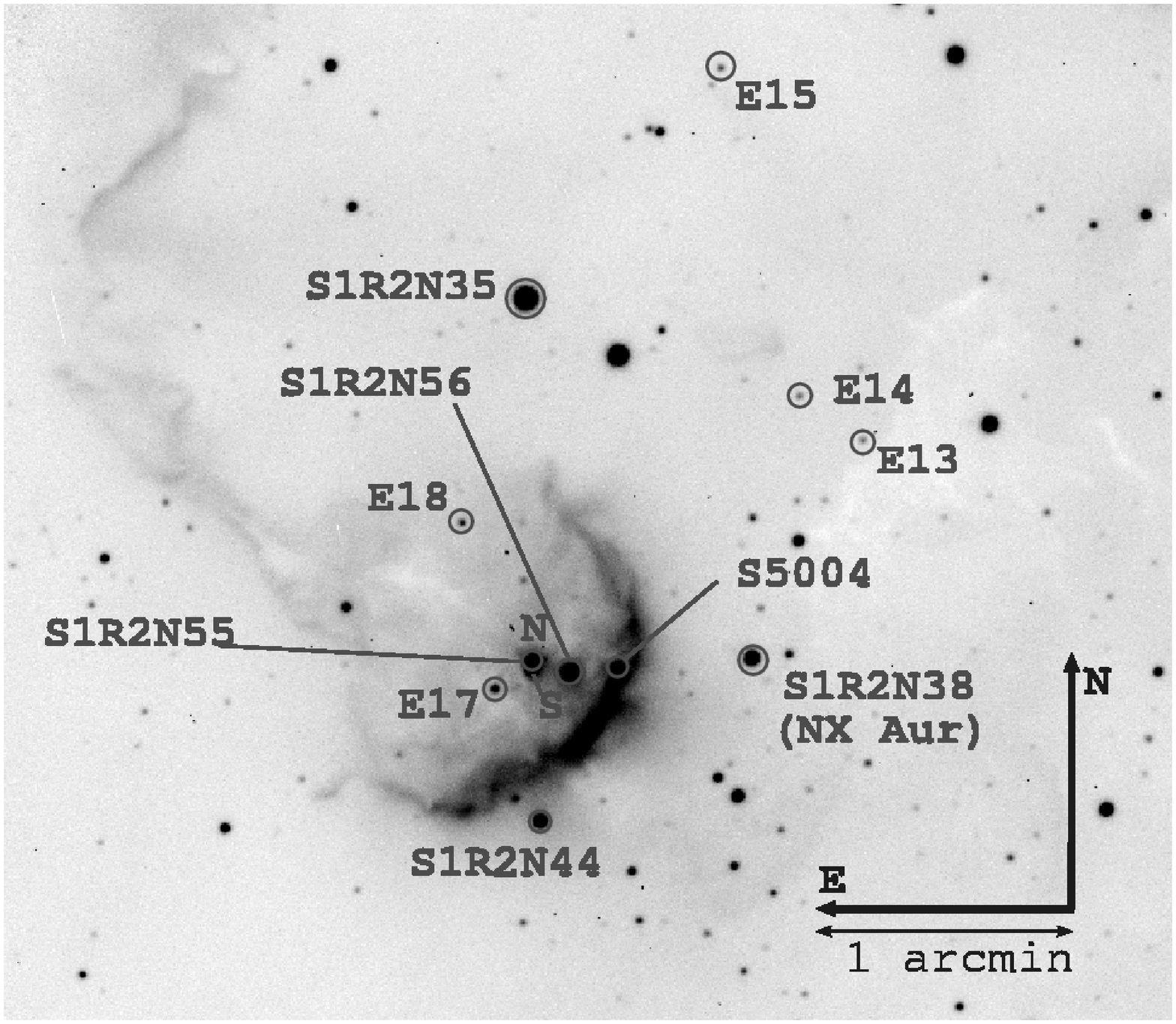}}
\caption{An H$\alpha$ image of the head of the cometary nebula
\object{Sim 130} (a portion of a 300-s exposure obtained on the night
of 5th December 2001 with ALFOSC on the NOT, equipped with the
narrow-band H$\alpha$ filter \#21).
The stars discussed in Section~\ref{sim} have been
identified. Note the bow-shaped emission rim and the extended
nebulosity surrounding the stars. } 
\label{fig:closeup}
\end{figure}

\subsection{S1R2N44}

S1R2N44 lies very close to Sim 130, just to the SW. Its
spectrum, displayed in Fig.~\ref{fig:normal}, shows nebular emission
lines on top of a normal absorption B-type stellar spectrum. A lower
resolution spectrum does not show any signs of intrinsic emission in
H$\alpha$. However, a cut of the spectrum in H$\alpha$ clearly shows
that the intensity of nebular H$\alpha$ 
emission increases considerably around the source, strongly suggesting
the association of S1R2N44 with the nebulosity. 

From the Full Width at Half Maximum (FWHM) of four \ion{He}{i} lines, 
we estimate for S1R2N44 an apparent rotational velocity
$v\sin i \approx 320\:{\rm km}\,{\rm s}^{-1}$, following the procedure
described by \citet{steele99}. We estimate its spectral type at  B2.5\,V.
For this star, \citet{fitz93} gives  $(b-y)=0.273$, implying
$E(b-y)=0.39$, slightly above the average for cluster members. If the
reddening is standard, $V=13.3$ implies $M_{V}=-1.9$, which is not
surprising for the spectral type. This star
must have reached the ZAMS, as its observed $(J-K_{\rm S})=0.25$ indicates
that this object has no infrared excess.

\subsection{S1R2N56}

This star forms the ``head'' of Sim 130. Its spectrum is strongly
contaminated by nebular emission, but this can be easily distinguished
from photospheric features at our resolution. There does
not seem to be emission intrinsic to the star (see the H$\alpha$
profile in Fig.~\ref{fig:red}, where also a weak
\ion{He}{i}~$\lambda$6678\AA\ absorption line can be seen).

 \begin{figure}
\begin{picture}(250,220)
\put(0,0){\includegraphics{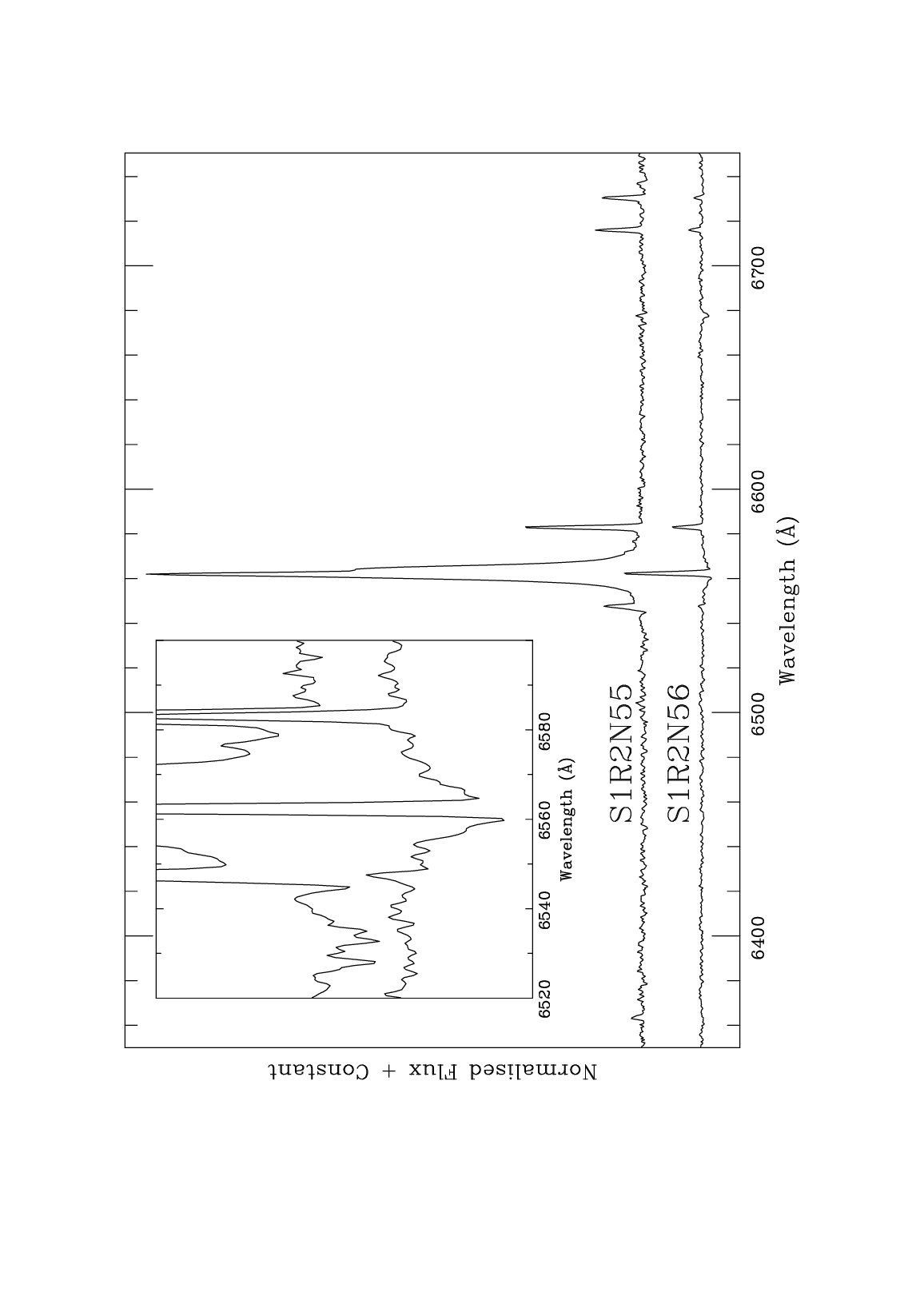}}
\end{picture}
\caption{H$\alpha$ spectra of the two stars in the
    ``head'' of Sim 130. While the emission features on the
spectrum of S1R2N56 seem to be entirely due to the
surrounding nebulosity, S1R2N55 is clearly an emission-line
star (see the inset for details). This object is the counterpart to
    the IRAS source 05198+3325, identified as a massive young stellar
    object.} 
\label{fig:red}
\end{figure}

The blue spectrum of this object is displayed in
Fig.~\ref{fig:normal}. From the FWHMs of
four \ion{He}{i} lines, we estimate an apparent rotational velocity
$v\sin i \approx 210\:{\rm km}\,{\rm s}^{-1}$. We estimate its
spectral type at B1.5\,V. For this object, \citet{mas95} measure
$(B-V)=0.43$, implying a 
reddening $E(B-V)=0.66$, well above the average for cluster
members. If the reddening law is standard, then the measured $V=13.54$
implies $M_{V}=-2.0$, which, though slightly too low for the
  spectral type, is not in strong disagreement. The 2MASS 
colour $(J-K_{\rm S})=0.76$ implies substantial reddening and
this object is selected as an infrared excess candidate, but this
could mainly be due to contamination of its photometry by the bright
nebulosity. 
 The available evidence suggests that S1R2N56 is a star
 settling on to the ZAMS.

\begin{figure}
\resizebox{\hsize}{!}{\includegraphics[bb= 40 145 340 695, angle=-90]{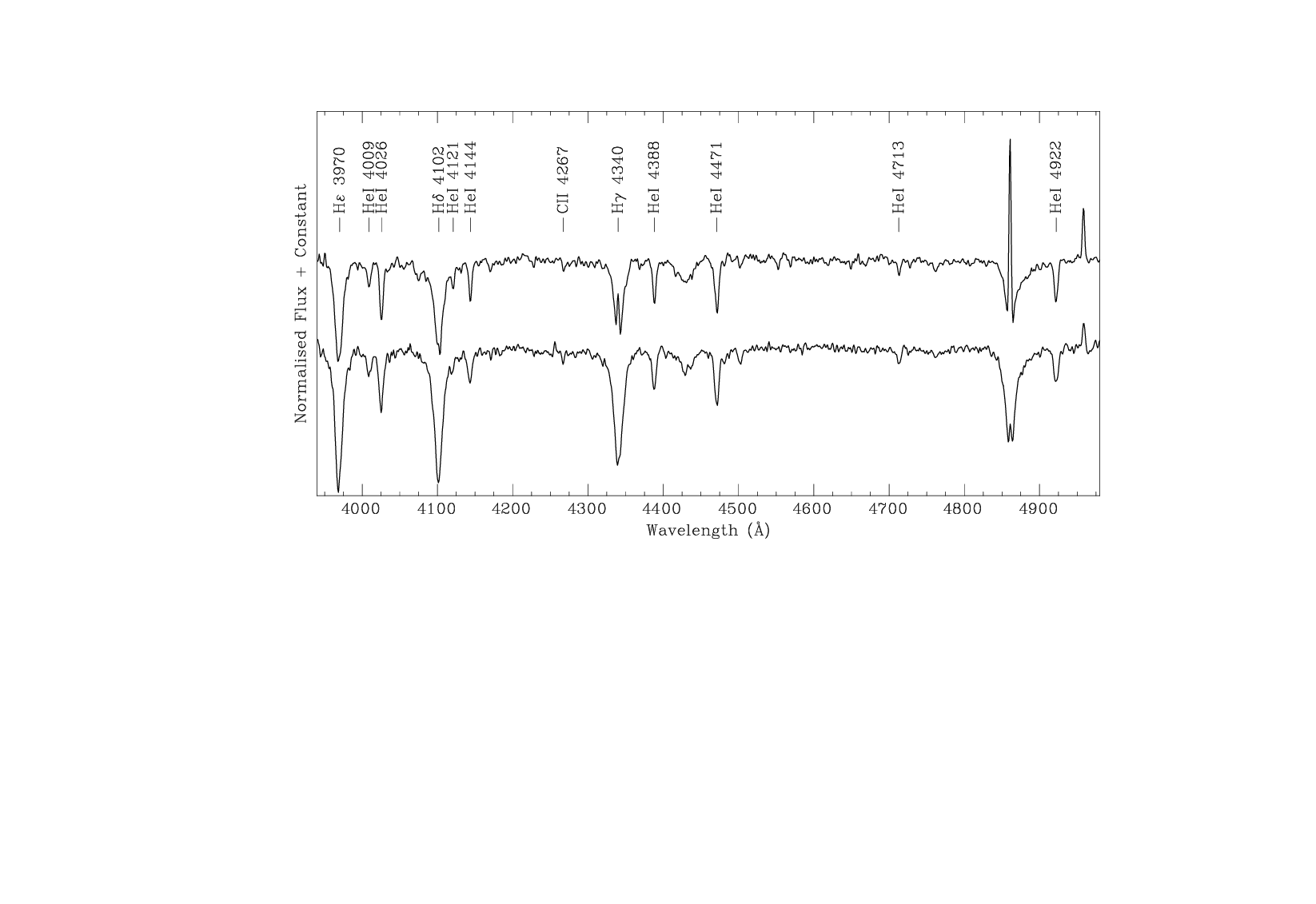}} 
\caption{Blue spectra of S1R2N56 (top) and S1R2N44. S1R2N56 is
    immersed in the pennant nebula Sim~130 and its spectrum shows
    strong nebular emission. Nebular emission is also present, though
    rather weaker, in the spectrum of S1R2N44, which lies just outside
    Sim~130. } 
\label{fig:normal}
\end{figure}

Observations
by previous authors seem to indicate very large variability in $V$
with a long-term dimming from $V<13$, but since none of these authors
mentions S1R2N55, it is possible that both stars have been
measured together (this is certainly the case in \citealt{tap91},
whose star \#35 corresponds to \object{S1R2N56}+\object{S1R2N55}).

\subsection{S1R2N55}

S1R2N55 is clearly immersed in the nebulosity associated with
Sim 130. As a matter of fact, our NOT  images clearly
show S1R2N55 to be a close double, but the two components
were not resolved during the OHP run, when spectra were taken.

The red spectrum is shown in Fig.~\ref{fig:red}. In 
addition to strong nebular lines, a very broad H$\alpha$
emission line can be 
seen: S1R2N55 is a Be star. The H$\alpha$ line peaks at 15
times the continuum intensity and has an Equivalent Width (EW) of
$-54\pm3$ \AA, a fraction of which is attributable to the nebular
component.  Given its presence in the middle of bright nebulosity in
an \ion{H}{ii} region with active star formation, S1R2N55 is
almost certainly a PMS Herbig Be star. In the $R$-band, the Northern
component is slightly brighter than the Southern one, but in the
H$\alpha$ images, it is much brighter. This clearly shows that the Northern
component of S1R2N55 is the emission-line object.

\citet{ish01} identified S1R2N55 as the counterpart
to the IRAS source \object{IRAS 05198+3325} (YSO CPM16). In their $K$-band
spectrum, they found strong Br$\gamma$ emission from the source and
{\sc H$_{2}$} emission of nebular origin.

\begin{figure}
\resizebox{\hsize}{!}{\includegraphics[bb= 40 145 430 695, angle=-90]{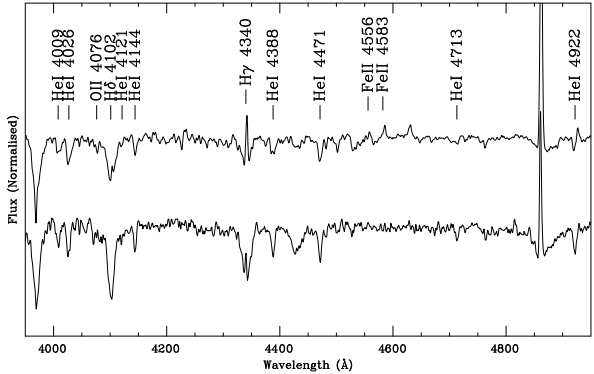}} 
\caption{Blue spectra S2R1N35 (top) and S1R2N55, two early Herbig Be
    stars associated with Sim~130. S1R2N55 is actually a
    blend of two very close B-type stars, and likely only one of them
    is a Herbig Be star.} 
\label{fig:herbigs}
\end{figure}

The blue spectrum of S1R2N55 is shown in
Fig.~\ref{fig:herbigs}. Emission is present in H$\beta$ and
H$\gamma$. The spectrum is likely dominated by the brighter Northern
component, but must be a combination of the spectra of the two
stars. From the apparent presence of weak \ion{O}{ii}~4076\AA\ and
\ion{Si}{iv}~4089\AA, the brighter component must be around  B1\,V. The
fainter component cannot have a very different spectral 
type, around B2\,V. This object
is selected among our infrared excess candidates. At least the bright
component is a Herbig Be star.

\subsection{S5004}

The faint star S5004 is located just on the bright rim of
nebulosity defining the head of the cometary nebula. In spite of
this, we have been able to clean the spectrum of nebular emission. The
spectrum shows a prominent G-band, comparable in intensity to
H$\gamma$, indicating a spectral type close to G0. The weakness of
the \ion{Sr}{ii}~4077\AA\ and other luminosity indicators seems
  compatible with
a G0\,V star. \citet{mas95} give $V=15.79$, $(B-V)=1.27$ for
this object. This star is prominent in the $K$-band ($K_{{\rm
    S}}=11.55$) and is selected as one of our infrared excess
candidates. Indeed, comparison of the observed $(J-K_{{\rm S}})= 1.40$
with the colours expected for this spectral type 
\citep{ducati} implies $E(J-K_{\rm S})=1.13$. This value is much
larger than would correspond to its $E(B-V)$, indicating a large
infrared excess, typical of a PMS star. If this is object is a cluster
member, then its intrinsic magnitude would be $M_{V}\approx-1.7$,
clearly very bright for a normal main-sequence G-type star. 
S5004 may then be an
intermediate mass star still on the contraction track, similar to
$\theta^{1}$ Ori E \citep{hg06}. A spectrum around the
\ion{Li}{i}~6707\AA\ line could test this hypothesis. Alternatively,
the photometry could be in error because of nebular contamination and
this could be a foreground star. 

\subsection{E17}

This star, too faint in the optical to have been observed by
\citet{mas95}, is very bright in the $K$-band ($K_{{\rm S}}=11.49$ in
2MASS) and is selected as an infrared excess object. Its
spectrum is completely featureless, except for a prominent H$\alpha$
emission line EW$=-40\pm2$\AA. The emission line and infrared
excess identify this object as a PMS star.

\subsection{S1R2N35 = E01}
Classified by \citet{kw99} as an emission line star, this object in the
immediate vicinity of \object{Sim 130} was
proposed as a Herbig Be candidate in \citetalias{chopi}, based on a
low-resolution spectrum. A higher resolution spectrum is shown in
Fig.~\ref{fig:herbigs}. Classification
is complicated by the strong emission lines, but, based on the presence of
some weak \ion{O}{i} lines and possible strength of \ion{Si}{iii}
lines, while the \ion{Si}{iv} lines are not
visible, we adopt a B1\,V, though it could be slightly later.

\citet{mas95} give $V=12.33$, $E(B-V)=0.52$, indicating a very large
colour excess $E(B-V)\simeq0.8$. Again, this object is selected as an
infrared excess candidate.  From the intrinsic colour of a B1\,V
star, the implied excess is $E(J-K)_{{\rm S}}=0.57$. This object is
hence a Herbig Be star. 

\subsection{S1R2N38 = E02}

Already known as an emission-line and variable star (NX~Aur), this
object lies in the immediate vicinity of Sim 130, displaying 
H$\alpha$ and H$\beta$ strongly in emission,
EW(H$\alpha$)$=-55\pm2$\AA\ and EW(H$\beta$)$=-3.7\pm0.3$\AA. The upper
Balmer lines display weak blue-shifted emission components. Prominent
emission lines of \ion{Fe}{ii} are also present. Though the object is
clearly a B-type star, an exact spectral type is difficult to
derive from our low-resolution spectrum. Based on the strength of the
\ion{Mg}{i}~4481\AA\ line, we estimate it to be B4\,V.

\citet{mas95} give $V=14.41$, $E(B-V)=0.57$, indicating a
$E(B-V)\approx0.8$. 2MASS gives $K_{\rm S}=10.25$ and $(J-K_{{\rm S}})=1.76$,
implying a huge infrared 
excess $E(J-K_{{\rm S}})\approx1.9$. This object is hence also a
Herbig Be star. 

\section{Discussion}
\label{disc}

\subsection{Distance, reddening and extent}
\label{sec:dist}

Our analysis shows that the optical/near-IR spectral energy
distributions of almost all stars earlier than B3 are best fit when a
standard $R=3.1$ reddening law is assumed. A standard interstellar law has
also been found by \citet{ys01}, using less sophisticated
techniques. \citet{tap91} found a value of $R=2.8\pm0.1$, slightly
lower than our value. On the other hand, all stars later than B3 show
evidence for an abnormal law, in the sense that $E(J-K_{\rm S})$ is
larger than expected from the measured $E(B-V)$. We interpret this
result as showing that stars later than B3 have infrared excess
emission, rather than non-standard reddening. This excess should be a
consequence of the presence of some remnant of the disk from which the
stars formed and points to a very young age for NGC~1893.

The presence of this emission excesses likely accounts for the
existence of some dispersion in the distance estimates for NGC~1893,
ranging from DM$=13.1$ \citep{tap91} to $13.9$ \citepalias{chopi}, as
the different procedures used to deredden the data will treat this
excess differently. In view of this difficulty, we prefer to assume
the distance derived from spectroscopic parallaxes using a standard
$R=3.1$ reddening law, i.e., a value $d \sim 5\:$kpc, as this is in
line with the distances to tracers of the Outer Arm in the
vicinity of NGC~1893 \citep{outer}. 

Since the area covered in this investigation is larger than the area
covered  by our photometry \citepalias{chopi}, we can use the photometry of
\citet{mas95} to identify B-type members outside the cluster core. In
the $B$/$(U-B)$ diagram, 
they form a very well defined sequence at bluer colours than any
other star in the field. Comparison to the estimated spectral types of
\citetalias{chopi} shows that B-type members are delimited by
$(U-B)<-0.1$ and $B<15.5$. This agrees well with the assumption of
standard reddening, as a B9 star has $(U-B)_{0}=-0.57$ and
therefore for the average reddening to the cluster
$E(U-B)=0.72E(B-V)=0.38$, a B9 star should have $(U-B)=-0.19$. 
Three foreground stars can be easily identified because of their
bright $B$ and position in the $V$/$(B-V)$ diagram, leaving $\sim60$
stars along the cluster sequence. Of them, 25 stars are
spectroscopically confirmed to be B2 or earlier: 5 O-type stars and 20 
in the B0-2 range (of which 4 show emission lines).

\subsection{Age}

Is there an age spread in NGC~1893? We do not find any evidence for a
deviation from the ZAMS for almost any member. The only star that
seems (slightly) 
evolved is the O7.5\,V S3R1N5. However, there are other possible
interpretations to its spectrum. For example, it could be the
composite of two O-type stars with slightly different spectral
types. We believe this 
to be rather possible, as this star lies in the densest region of the
cluster and its spectroscopic distance modulus is rather shorter than
the average for the cluster. Moreover, it is surrounded by PMS stars
and shows a strong IR 
excess. Therefore we do
not think that this is an evolved star, especially as nearby stars of
earlier spectral type do not show any sign of evolution. The fact that
an O4\,V and an O5.5\,V star are still close to the ZAMS places a
strict upper limit on the age of the cluster at 3~Myr and supports an
even younger age \citep{mm03}. Comparison with PMS isochrones
(Fig.~\ref{fig:excess}) suggests an age of $\la 2$~Myr, in good
agreement  with the fact that stars 
as massive as B3\,--\,4\,V still show significant $E(J-K_{\rm S})$
excesses, likely 
due to the presence of remnants of the disks from which they formed.

However, it is obvious that, while an important population of B-type
stars, some of them as late as B8\,--\,A0, is already settling into the
ZAMS, some massive stars, with spectral types in the B1\,--\,B2 range are
still in the Herbig Be phase. This shows that there is some spread in
the formation of stars. In this paper, we have selected the PMS stars
with the strongest signatures of youth (emission lines and/or strong
$E(H-K_{\rm S})$ excess) and found that their distribution is limited
to two small regions within the relatively large area covered by MS
members. The fact that recent star formation is confined to these two
regions suggests that we are observing the star formation process
spreading from the central cluster to the neighbouring dark cloud. 

The lack of three-dimensional information does not allow us to
determine the relationship of the young PMS stars close to the cluster
centre with the members already on the main sequence. Images of the
area suggest that part of the dark cloud is partially hiding the
cluster and perhaps most of the star formation is taking place on the
inner wall of the cloud, which we cannot see. However, it seems that
the population of PMS stars to the East of the cluster is being formed
on the illuminated surface of the molecular cloud around the two
bright pennant nebulae Sim~129 and Sim~130.

\subsection{Triggered star formation in Sim~130}

As described in Section~\ref{sim}, the area surrounding the pennant
nebula Sim~130 contains three Herbig Be stars and several other
emission-line stars. Other emission-line stars cover the area between
Sim~130 and Sim~129. In the vicinity of Sim~129, there are two early
A-type PMS stars, S1R2N4 \citepalias{chopi2} and S1R2N26 (Table~2). 

The impact of the ionising flux from the O-type stars on the nebulae
is obvious. Their cometary aspect is due to the presence of bright
ionised fronts, taking a shape strongly resemblant of a bow shock,
combined with a ``tail'' that seems to run away from the
centre of the cluster and is actually composed of bright filaments
illuminated by the O-type stars. Moreover, similarly to other star forming
regions in the vicinity of massive clusters \citep[e.g.,
  M16;][]{hester}, both nebulae show finger-like dust structures oriented
towards the nearby O-type stars. 

In view of these properties, the area of recent star formation
around Sim~130 presents all the
characteristics listed by \citet{wal02} as typical of
regions of triggered star formation:
\begin{itemize}
\item The younger (second generation) stars are associated with dust
  pillars oriented towards the O-type stars.
\item The second generation is less massive than the first. In this
  case, we have 6-7 early and mid B-type stars, as compared to the
  $\sim20$ massive stars in the main cluster.
\item The more massive stars in the second generation are less massive
  than the more massive stars in the first cluster (in this case, the
  earliest spectral type around Sim~130 is B1\,V, as opposed to O4\,V
  in the main cluster).
\end{itemize}

\citet{wal02} finds a characteristic age difference of $\sim$2~Myr
between the first and second generation. This may indeed reflect the
smallest age difference that we are able to detect, but the observed
properties of the two populations in NGC~1893 are not incompatible
with such an age difference.
 
Obviously, it may be argued that the stars around Sim~130 may
represent star formation that would have occurred regardless of the
effect of the nearby massive stars and that it is simply being exposed
now because of the erosion of the molecular cloud by the ionising flux of the
O-type stars. Even though it is difficult to find strong arguments
against this view, it should be noted that {\it MSX} images of the
molecular cloud associated with IC~410 do not give any reason to
suggest that star formation is taking place anywhere except in the two
areas marked by emission-line stars, just over the surface of the
molecular cloud. It would be surprising if the only places in the
molecular cloud in which star formation is spontaneously occurring
happen to be, just by chance, in the process of being cleared up by
the ionising flux of the nearby O-type stars. 

%Recent theoretical models showing how the presence of massive
%($M_{*}\ga 20M_{\sun}$) stars may give rise to a triggering expanding
%shell have been presented by \citet{hi06a,hi06b}.

\subsection{The PMS stars}

We have shown that, with the possible exception of the emission-line
star E09, there are no obscured massive stars. We can consider the
census of stars earlier than B3 complete. E09 is very bright in
$K_{{\rm S}}$ and extremely reddened. It is likely associated with the
IRAS source 05194+3322\fnmsep\footnote{This source is identified in SIMBAD
  with the O7.5\,V member HD~242935, but its coordinates coincide much
better with E09.} and therefore may be a massive young stellar
object. 

We cannot define a strict detection limit for emission-line stars, but
it is very unlikely that the slitless observations may have skipped
any relatively bright ($V<15$) stars. This means that, while there are
4 early Herbig Be stars in NGC~1893 (see Table~\ref{tabber:new}),
there is one mid Herbig Be star, no late Herbig Be stars and only one 
Herbig Ae star (also, as can be seen in Fig.~\ref{fig:(J-K)-K}, the
$K$ magnitudes of all 
the candidate infrared excess stars are too faint to expect any of
them to be B-type stars). This fact is difficult to
interpret. Obviously, any reasonable 
IMF should result in a rather larger number of intermediate-mass stars
than massive stars. Moreover, the more massive the star is, the
earlier it should reach the ZAMS. Finally, the UV flux of the early
B-type stars should help to dissolve their disks much more quickly
than those around lower mass stars. 

How can we then explain the excess of early Herbig Be stars with
respect to other emission-line stars? The possibility that early
Herbig Be stars retain their disks longer than later type Herbig ABe
stars is counterintuitive. If the star formation process in
the two active areas is very recent, it may be possible that
only the early Herbig Be stars have emerged from the parental cloud
and less massive stars are too faint to be observed in the
optical. In this view, the mid and late B stars that we see are
associated with the first generation of massive stars, while the
second generation is now emerging from the parental cloud, likely
because the UV photons from the O-type stars are photodissociating the
cloud around them. The later-type emission line stars should then also
be associated with the first generation of massive stars.

\subsection{The IMF}

The extreme youth of NGC~1893 offers a good prospect for determining
the IMF of a population just emerging from the parental
cloud. For this, deep infrared observations would be needed in order
to probe the low-mass stellar populations. However, some difficulties
stand out.

First, as discussed above, it is possible that part of the cluster is
obscured by parts of the dark cloud. Assuming typical masses for
spectral types, the observed distribution of
members is 5 stars with $25M_{\sun}\leq M_{*}\leq60M_{\sun}$ (O4\,--\,07.5),
20 stars with  $8M_{\sun}\leq M_{*}\leq16M_{\sun}$ (B0\,--\,B2) and $\sim35$
stars with $3M_{\sun}\leq M_{*}\leq8M_{\sun}$ (B2.5\,--\,B9). This
distribution seems too biased towards early spectral types for a
normal IMF. For a Salpeter IMF, which is valid in this mass range
\citep{kroupa}, we would expect three times more intermediate mass 
(B2.5\,--\,B9) than massive($\leq$B2) stars. 

 Ignoring any incompleteness due to multiplicity, the 25 massive stars
 with known spectral types result in a mass $\sim 400M_{\sun}$. Assuming
 for the sake of argument that the deficit in 
intermediate mass stars is due to observational effects and the IMF is
standard \citep[i.e.,][]{kroupa}, this mass implies a total mass
$M_{{\rm cl}}\sim 2200M_{\sun}$ for NGC~1893. This estimate is a lower
 limit, as there are reasons to believe that a substantial fraction of
 the most massive stars in the cluster are, at least, binaries: a few
 radial velocity measurements by \citet{jon72} show most of them to
 display large velocity variations.  

A second factor suggesting obscuration of part of the cluster is its
shape. The distribution of members is traced in
Fig.~\ref{fig:ms_pmsdistribution}. If we take the 
intermediate mass stars as best tracers of the cluster extent, it is
difficult to assign a morphological type or even define a centre. The
main concentration  of stars appears just on the edge of the molecular
cloud. The conspicuous absence of any likely member to the West
  of the cluster core strongly suggests that NGC~1893 is located on
  the back side of the molecular cloud associated with IC~410.

An even more striking difficulty is the fact, evident in
Fig.~\ref{fig:ms_pmsdistribution}, that 
there is a halo of high-mass stars surrounding the cluster in areas
where there are essentially {\it no} intermediate-mass members. This
is more obvious to the East of the cluster, beyond Sim~130. If we
consider the area lying between S2R3N35 (RA: $05^{h} 23^{m} 13^{s}$))
and the edge of Figure~\ref{fig:ms_pmsdistribution}, it contains the 8 members 
identified in Fig.~\ref{fig:genview}. These objects lie at distances of
$8\arcmin$ to $12\arcmin$ from the cluster core (corresponding to 12
to 18 pc at $d\sim5\:$kpc). As seen in
Table~3, 7 of them have spectral types in the
B0\,--\,1 range and one is a mid B-type star. This area is fully covered by
the photometry of \citet{mas95}, which provides only three other
  photometric members. The brightest one, [MJD95] J052325.13+332609.8
  was classified as 
B1.5 by \citet{mas95}, while the two other photometric members,
[MJD95] J052336.60+332905.5 and J052339.25+333839.7 have colours and
magnitudes appropriate for mid-B stars (note that J052339.25+333839.7
falls outside the area covered in this investigation and lies outside
Fig.~\ref{fig:ms_pmsdistribution}, $\sim 4\arcmin$ to the North of the
Northernmost members displayed). Therefore, this area includes 8
early B stars, 3 mid B stars and no
late B-type photometric members. How do we arrive at such surprising mass
distribution? 
 
An obvious candidate for an explanation is dynamical ejection from the
cluster core. As discussed by \citet{ld90}, massive clusters
containing hard binaries with two components of similar mass may be
quite effective at ejecting stars via dynamical ejections. The
majority of stars ejected will be B-type stars and their ejection
velocity will be inversely proportional to their mass. These factors
could explain a concentration of early B-type stars in the vicinity of
a young massive cluster, but the efficiency at ejecting stars
estimated by \citet{ld90} seems much lower than that required to
explain the population of objects around NGC~1893. However, more
recent work by \citet{pflam} suggests that, if massive stars are
mainly born as part of multiple systems, the ejection rates can be
much higher than estimated by \citet{ld90}. According to their
results, a cluster with a mass comparable to that 
of NGC~1893 could lose $\sim 75\%$ of its high mass stars in
1\,--\,2~Myr. 

In this respect, it is interesting to note that there is one
  further early-type star, LS V +33$\degr$31, classified B0.5\,V by 
\citet{mas95}, lying about $24\arcmin$ away from the cluster core,
  which falls along the sequence of cluster members. In the whole area
  covered by \citet{mas95}, there are only three more objects that
  could be photometric members, all very distant from the cluster
  ($d>15\arcmin$) and all compatible with being late B-type stars
  [MJD95] J052425.46+331544.7, [MJD95] 
J052346.82+331440.6 and [MJD95] J052516.96+332403.9.
Even more striking is the fact that HD~242908, nominally the most
massive star in the cluster, lies at some distance from the main bulk
of the cluster, in an area where very few other members are
found. Intriguingly, if we do not count stars in this massive star
halo, the ratio between massive and intermediate-mass stars is close
to standard. 

Radial velocity measurements of the stars in the halo of NGC~1893
could provide a test on this hypothesis, as the systemic velocity of
the cluster is close to zero \citep[e.g.,][]{jon72} and any measured
components should be due to runaway velocities. 

One further issue to take into account is the complication arising
from the presence of sequential star formation. If sites of triggered
star formation (such as Sim~130) contribute only stars less
  massive than $\sim12M_{*}$ (B1\,V), the total integrated
  (first generation + triggered generations) population will have a
steeper IMF than 
the original first generation. If we observe this cluster in a few
Myr, there will be no way of telling which stars have formed at which
time. Of course, this has a bearing on how we can define an instantaneous
IMF.

\section{Conclusions}
 
   \begin{enumerate}
\item We have found a population of emission-line PMS stars in
  NGC~1893. The brightest among them cover the range from B1 to late
  F, with an obvious overpopulation of early B-type stars. Emission
  line stars appear only in two regions of the cluster.
\item We have identified a number of faint objects with high values of
  $(J-K_{{\rm S}})$ that seem to show an infrared excess. These objects
  concentrate around the emission-line stars, indicating that they are
  also PMS stars.
\item All the stars later than B3 show evidence for an infrared
  excess, even though the main sequence is well traced down to A0 in
  the optical. This infrared excess increases as we move to
  later spectral types, strongly suggesting that it arises from the
  remnant of a disk. 
\item The age of NGC~1893 is constrained to be $<3$~Myr by the
  presence of main-sequence 04 and O5 stars and likely to be
  $\la2$~Myr. This makes NGC~1893 one of the youngest clusters to be
  visible in the optical. It is very likely in the process of emerging
  from its parental cloud and perhaps more members lie hidden by
  dark portions of the cloud. If this is the case, they are quite
  faint and infrared observations reaching deeper than 2MASS are
  needed to detect them. 
\item The area around the cometary nebulae Sim~129 and Sim~130 shows
  the highest number of emission-line and IR-excess PMS stars. Three
  B1-B4 Herbig Be stars cluster around Sim~130. This is likely to be a
  region of more recent star formation, triggered by the ionisation
  front generated by the O-type stars.
\item A second region containing emission-line stars and IR-excess PMS
  candidates lies on the interface between the cluster core and the
  molecular cloud. Here we could have another area of triggered star
  formation partially hidden by the molecular cloud. On the very edge
  of the cloud, we find the emission line object E09, which, with
  $K_{{\rm S}}=9.4$ and $(J-K_{{\rm S}})=3.0$, could be a massive very
  young stellar object. 
\item The picture of star formation emerging from our study of
  NGC~1893 is a rather complex one, with sequential star formation
  resulting in several slightly non-coeval populations and dynamical
  ejection depopulating the cluster of massive stars at a very young
  age.  
\end{enumerate}

\begin{acknowledgements}

IN is a researcher of the programme {\em Ram\'on y Cajal}, funded by
the Spanish Ministerio de 
Ciencia y Tecnolog\'{\i}a (currently Ministerio de Educaci\'on y
Ciencia) and the University of Alicante, with partial 
support from the Generalitat Valenciana and the European Regional
Development Fund (ERDF/FEDER).
This research is partially supported by the MEC under
grant AYA2005-00095 and by the Generalitat Valenciana under grant
GV04B/729.    

During part of this work IN and AM were visiting
fellows at the Open University, whose kind hospitality is warmly
acknowledged. IN was funded by the MEC under grant
PR2006-0310. AM was funded by the Generalitat Valenciana
under grant AEST06/077.

 The INT is operated on the island of La
Palma by the Isaac Newton Group in the Spanish Observatorio del Roque
de los Muchachos of the Instituto de Astrof\'{\i}sica de Canarias. 
Based in part on observations made at 
Observatoire de Haute Provence (CNRS), France. IN would like to
express his thanks to the staff at OHP for their 
kind help during the observing run.
The Nordic Optical Telescope is operated on the island of La Palma 
jointly by Denmark, Finland, Iceland, Norway, and Sweden, in the
Spanish Observatorio del Roque de los Muchachos of the 
Instituto de Astrofisica de Canarias. Part of the data presented here have 
been taken using ALFOSC, which is owned by the Instituto de Astrof\'{\i}sica 
de Andaluc\'{\i}a (IAA) and operated at the Nordic Optical Telescope under 
agreement between IAA and the NBIfAFG of the Astronomical
Observatory of Copenhagen.
 
This research has made use of the Simbad data base, operated at CDS,
Strasbourg (France) and of the WEBDA database, operated at the
Institute for Astronomy of the University of Vienna. This publication
makes use of data products from 
the Two Micron All 
Sky Survey, which is a joint project of the University of
Massachusetts and the Infrared Processing and Analysis
Center/California Institute of Technology, funded by the National
Aeronautics and Space Administration and the National Science
Foundation.

We thank the anonymous referee for insightful comments that
helped us clarify some topics.

\end{acknowledgements}

\end{document}